**Green synthesis of nanomaterials from sustainable materials for biosensors and drug delivery.**

**Naumih M. Noah[1]* and Peter M. Ndangili[2]**

[1]School of Pharmacy and Health Sciences, United States International University-Africa (USIU-A) P.O Box 14634-00800, Nairobi, Kenya

[2]School of Chemistry and Material Science (SCMS), Technical University of Kenya P.O. Box 52428 – 00200 Nairobi- Kenya

*Corresponding Author; Email: mnoah@usiu.ac.ke; Phone: +254203606726, +254730116726

**Abstract**

The integration of biosensing platforms with drug delivery systems has led to effective treatment strategies for biomedical applications. With the emergency of nanotechnology, the manipulation of materials in the nanometer (nm) scale of 1 – 100 nm, these biosensing and drug delivery systems have been tremendously improved due to the exceptional properties exhibited by these materials. The conventional approaches used to synthesize the nanomaterials including physical and chemical methods involve the usage of harsh chemicals and hazardous reaction conditions and hence posing a threat to health and the environment. This problem is solved by the biological methods that involve green nanotechnology which integrates green chemistry and engineering principles to formulate harmless and eco-friendly nanomaterials to fight the complications affecting human health and the environment. These biological methods use phytochemicals found in plants and plants parts as well as microorganisms for the bioreduction of metal ions to their corresponding nanomaterials. The plants and the microorganisms are readily available, cost-efficient, and have biocompatibility hence offering sustainable synthetic methods for nanomaterials. This review will therefore focus on the plant-mediated and microbe-mediated green synthesis of different nanomaterials, the mechanisms of these synthetic methods, the application of the green synthesized nanomaterials in biosensing and drug delivery as well as the challenges of the synthetic methods.

## 1. Introduction

The detection of diseases at an early onset can enhance the efficacy of therapies including rapid diagnosis and treatment [1]. For improved health management, biosensors, which are devices that utilize the biological recognition components for the detection of disease biomarkers followed by the transduction of the biomarkers, have been used [2]. The latest improvements in biosensor technology have led to the development of accurate, precise, robust, and rapid systems that can sense disease-dependent changes in the analyte hence enabling a quick diagnosis and treatment of many diseases [1]. Integrating biosensing platforms with drug delivery systems has led to even more effective treatment strategies for biomedical applications such as in the treatment of diabetes [1].

Nanotechnology, which involves the manipulation of materials in the range of 1 to 100 nanometers has introduced new avenues for drug delivery and biosensing. Several nanotechnology-driven strategies have empowered researchers to fabricate various drug delivery and biosensing platforms [3]. For example, the use of nanomaterials in the development of biosensors has led to improvement of their sensitivity and performance allowing the introduction of several signal transduction technologies [4]. Nanotechnology has also made it possible to develop a nanoscale drug delivery system for different drugs to the target location.

Different approaches have been engaged in the synthesis of nanomaterials based on the nature and type of the nanomaterials. Generally, two strategies, namely “top-down” and “bottom-up” have been utilized. In the top-down strategy, bulk materials are reduced to nanomaterials while in the bottom-up strategy, the nanomaterials are synthesized from elementary materials [5]. These strategies have been used in synthesis methods such as the physical, chemical, and biological methods. The specific physical methods are; lithography, pyrolysis, physical vapour deposition, crushing, grinding, attrition, and ball milling. The chemical methods include chemical vapor deposition, hydrothermal, solvothermal method, sol-gel method, thermal decomposition, microwave-assisted synthesis, ultra-sonic assisted, reduction via photo catalysis, electrochemical and gas-phase among others [5]. These methods are characterized by uniform size distribution, homogeneity, and low energy consumption during synthesis.

However, they are associated with high energy demands in production, high costs, and low yields [6]. They involve laborious synthesis procedures, some chemicals used are cytotoxic, genotoxic, and carcinogenic, leading to the production of unsafe nanoparticles and environmental pollution [7], [8]. The toxic nature of chemically synthesized nanomaterials, compounded with their inherent instability makes them not economically feasible and environment-friendly [9], and this largely limits their biomedical applications [10]. This has led to the growth of economically feasible and environmentally synthetic strategies such as green nanotechnology.

Green nanotechnology offers tools for the conversion of biological systems to green methodologies to nanomaterial synthesis providing an impeccable solution to diminish the negative effects of the chemical and physical methods and application of nanomaterials preventing any accompanying toxicity hence lowering the nanotechnology riskiness [11][12]. It involves the integration of the principles of green chemistry and engineering outlined in table 1 to produce safe and eco-friendly nanomaterials that maximize their safety and sustainability [9], [11]. The current green synthetic methods reported use plant or plant parts as well as microorganisms such as bacteria, algae, fungi, and yeast among others for the bioreduction of metal ions into their corresponding nanomaterials [13] as illustrated in figure 1. These green methods are eco-friendly, cheap, and chemical contaminant free which is important in the application in the biomedical field where the purity of the nanomaterials is of concern [13][14]. This biogenic reduction is a “Bottom-Up” method where natural extracts from the plants or the microbes provide the reducing agents, stabilizing agents, growth terminating, and capping properties as well as influencing the size and shape of the nanomaterials [13]. The biomolecules, cells, and organs of the above-mentioned plants and microorganisms have been engineered to deliver innovative nanomaterials with potential sustainable advantages [12]. In this review, we, therefore, focus on the green synthesis of various nanomaterials using plant materials and microorganism and their applications in biosensing and drug delivery.

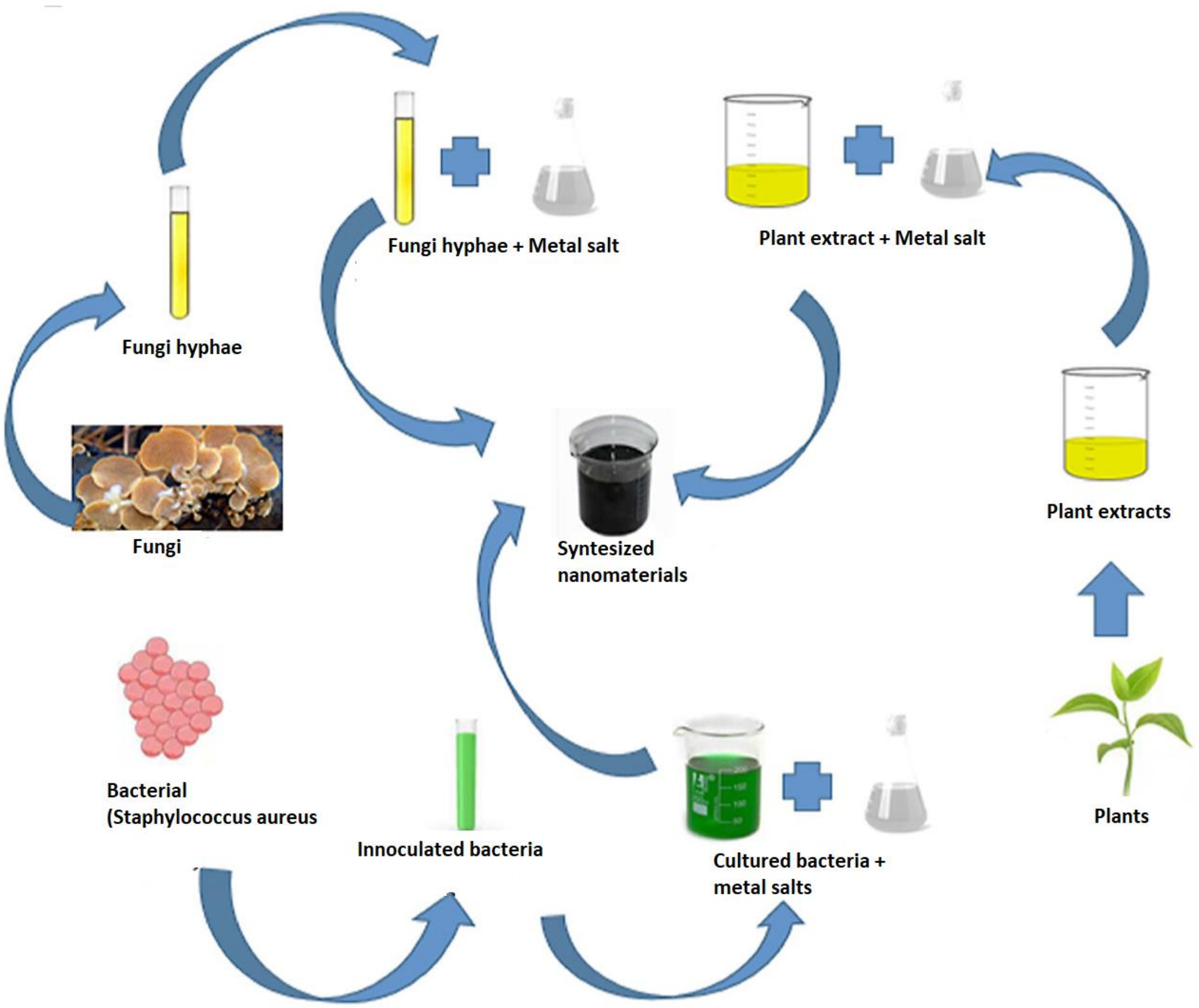

Figure 1: Schematic diagram showing the different biosynthetic methods used in the synthesis of nanomaterials

**Table 1: The principles of green chemistry and engineering in the design of greener nanomaterials**

| **Green Chemistry Principles in the design of greener nanomaterials** [9][15]**.** | **Green engineering principles in the design of greener nanomaterials** [16]**.** |
|---|---|
| **Prevention**: design of nanomaterials for waste reduction by determining their biological impact based on their measurable properties | **Inherent Rather Than Circumstantial:** the synthesis of nanomaterials should ensure that the raw materials used are kind and neutral to the environment. |
| **Atom Economy**: use synthetic methods that maximize efficiency by incorporating all materials used in the process into the final product | **Prevention Instead of Treatment:** the synthesis of nanomaterials should ensure that the inputs are designed as part of the outputs. |
| **Less Hazardous Chemical Syntheses:** synthetic methods that utilize and generate nontoxic substances to ensure process safety as well as degradable nanomaterials | **Design for Separation:** ensuring that the design of the nanomaterials contains essential chemical properties of volatility and solubility for the separation and purification of the raw materials in an environmentally friendly way. |
| **Designing Safer Chemicals:** synthesis of nontoxic nanomaterials to ensure safety. | **Maximize Efficiency**: adopting optimization strategies such as replacing batch or vessels reactors by spinning disk reactors or replacing paints with powder coatings to ensure efficiency. |
| **Safer Solvents and Auxiliaries:** avoiding the usage of auxiliary substances (e.g., solvents, separation agents) to ensure the safety of nanomaterials | **Output-Pulled Versus Input-Pushed:** adopting the "Push vs Pull" strategy in the design of nanomaterials to ensure that the end-user of the nanomaterials is also the final purchaser. |
| **Design for Energy Efficiency:** carrying out chemical reactions for the synthesis of nanomaterials at ambient temperature and pressure to ensure efficiency. | **Conserve Complexity**: taking advantage of the benefits from complexities involved in the design of the nanomaterials. |
| **Use of Renewable Feedstock: use of renewable** raw material or feedstock in the synthesis of nanomaterials to reduce environmental impact. | **Durability Rather Than Immortality**: designing nanomaterials with a goal of their durability instead of immortality in mind to minimize the threat and risks to humans and the environment. |
| **Reduce Derivatives:** Avoid unnecessary derivatization | **Meet Need, Minimize Excess**: the synthesis of nanomaterials should incorporate design ideas to reduce the expenditures for the resources. |
| **Catalysis: use of c**atalytic reagents in the synthesis of nanomaterials to ensure efficiency | **Minimize Material Diversity**: considering the end-of-life criteria for the nature, recycle or reuse of the nanomaterials by adopting easy to assemble and disassemble design ideas. |
| **Design for Degradation: using degradable** Chemical products in the synthesis of nanomaterials to reduce environmental impact. | **Integrate Material and Energy Flows**: by storing the byproducts obtained from the nanomaterials synthesis reactions. |
| **Analyze in real-time to prevent pollution**: Include in-process real-time monitoring and control during syntheses to minimize or eliminate byproducts | **Design for Commercial "Afterlife":** by developing synthesis ideas that use the existing products in the upcoming nanomaterials rather than investing new expensive resources. |
| **Minimize accidents**: Design nanomaterials and their forms to minimize the potential for chemical accidents | Renewable Rather than Depleting: by using biological renewable resources in the synthesis of nanomaterials. |

## 2. Plant Mediated Biosynthesis of Nanomaterials

Plants materials have been reported to have extraordinary bioreduction capability to synthesize numerous nanomaterials. This is because they contain rich natural compounds such as alkaloids, flavones, flavonoids, polyphenols, terpenoids, saponins, steroids, tannins, proteins, and other nutritional compounds derived from the various parts of the plant such as the bark, leaves, fruits, stems, roots, and seeds [17]–[19]. The plant extracts including leaves, roots, flowers, and barks are said to contain many secondary metabolites which act as reducing, stabilizing, and capping agents for the bioreduction reactions in the synthesis of nanomaterials [20][21][19] as illustrated in figure 2. The polyphenols have been reported to have the most prominent effect in both reductive and protective capability among all components [22] since they guard the plants against the reactive oxygen species (ROS) produced during photosynthesis and contact with anthropogenic impurities [12]. The plants can be used either in live form or inactive form to reduce the metal precursors to nanomaterials intra-cellularly [21], extracellularly or phytochemical mediated synthesis [17]. The concentration of polyphenols is critical in the preparation of nanoparticles especially AuNPs with the synergistic competition between the phenolic hydroxyl and carbonyl groups causing the oxidation of the polyphenols hence affecting the particle size and morphology of the nanoparticles [22].

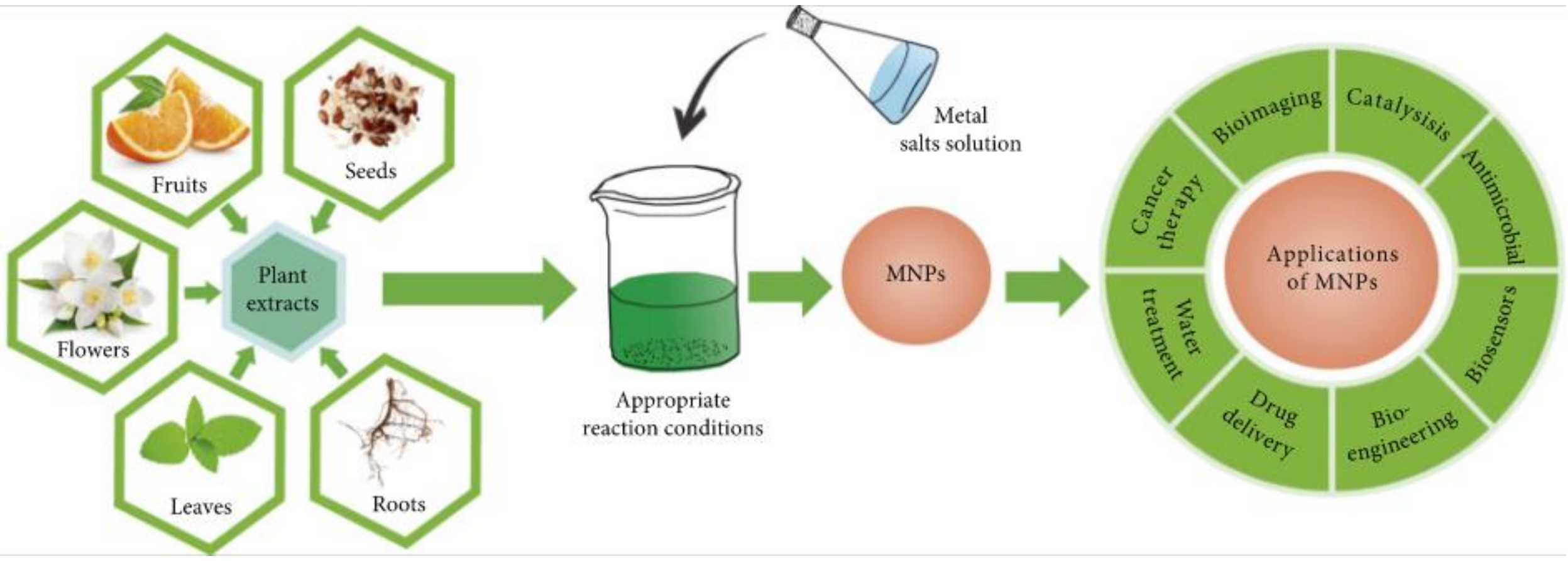

Figure 2: A schematic diagram representing the green synthesis of nanomaterials and their application. Reproduced from [23], an open-access article.

### 2.1: Mechanism of the plant-mediated synthesis of nanomaterials

The mechanism of the plant-mediated synthesis of nanomaterials is reported to take place in three phases, namely the reduction, the growth, and the termination phases [23]. The reduction phase involves the reduction of the metal ions to zero-valent metal atoms by reducing phytoactive molecules via electron transfer. Consequently, in the growth phase, the zero-valent metal atoms aggregate into nano-metallic particles with different shapes such as linear shape, rod-like shape, triangular, hexagonal, or cubic. Lastly, in the termination phase, the phytoactive constituents with antioxidant properties get enriched around the nanomaterials to maintain their stability [23]–[25]. This plant-mediated synthesis has known advantages such as the use of aqueous solvents, the plant materials being readily available, cost-efficient, the biocompatibility of the plant extracts [17] as well as the capability to support large scale synthesis [18]. The plant extracts also possess antioxidants which act as reducing agents for the nanomaterials while acting as a stabilizer to protect them from oxidation [26]. The mechanism of the plant-mediated synthesis of nanoparticles is illustrated in figure 3[27]. Most of the plants are sustainable and are featured as renewable suppliers of biomolecules used in the synthesis of nanomaterials as compared with microbes and enzymes, since they can pick up almost 75% of the light energy and transform it into chemical energy, with antioxidants and sugars, for the synthesis of nanomaterials [12]. Due to these advantages, the plant-mediated synthesis has been employed in the synthesis of various nanomaterials such as silver nanoparticles (AgNPs), gold nanoparticles (AuNPs), graphene, palladium nanoparticles among others which have found applications in numerous fields such as medicine, biosensing, wastewater treatment, etc [17]. In the next sections, we highlight the recent use of plant extracts in the synthesis of metal nanoparticles and metal oxide nanoparticles

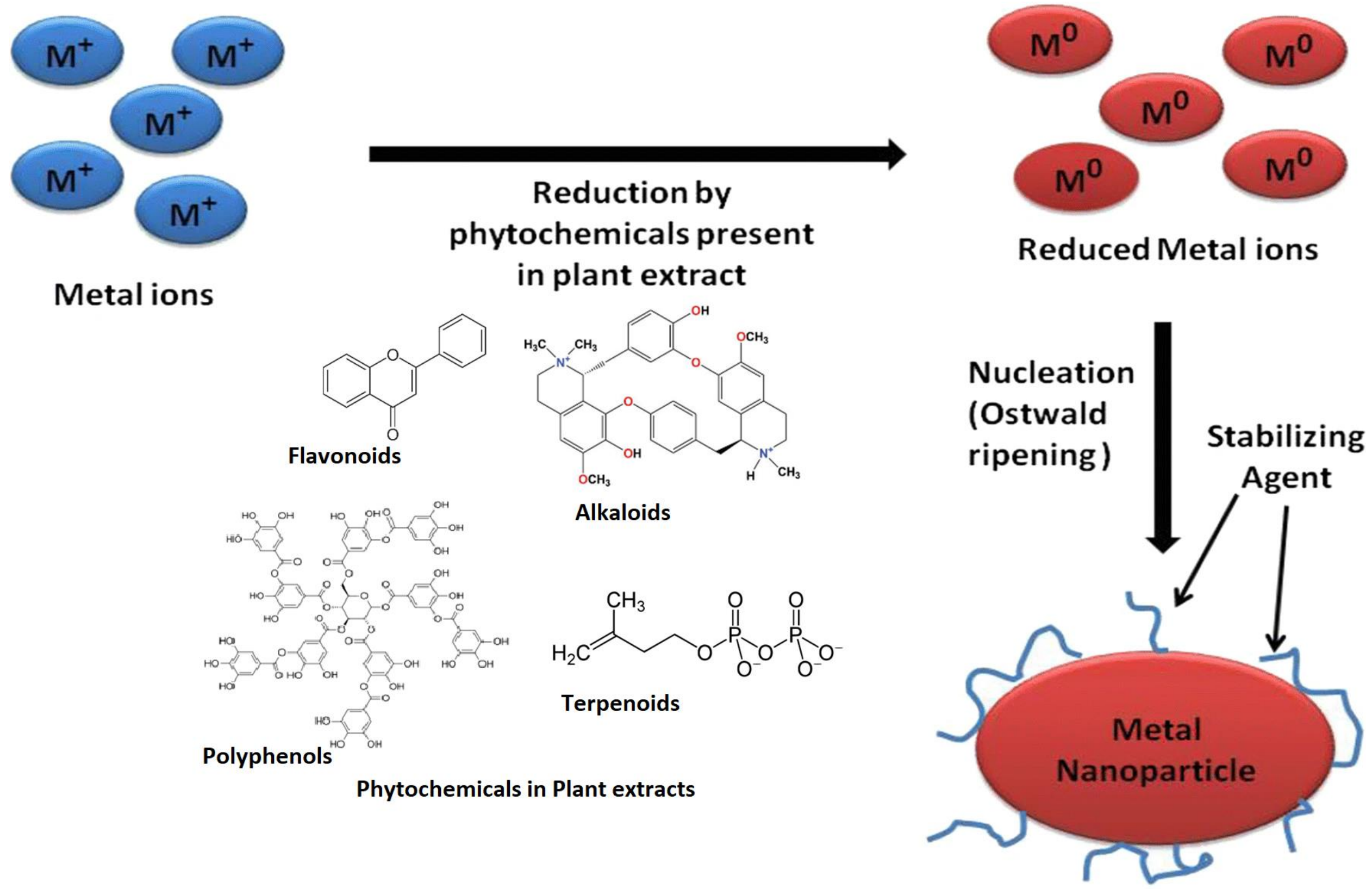

Figure 3: A schematic diagram representing the mechanism of the plant-mediated synthesis of nanoparticles. Reproduced with slight modification from [27], an open-access article.

**2.2: Plant-mediated synthesis of Silver Nanoparticles (AgNPs)**

Silver nanoparticles are among the most widely applied nanoparticles in the biomedical field, food production, and drug delivery owing to their excellent chemical stability, antimicrobial activity, catalytic properties, and electrical conductivity [23]. Their wide application has also been linked to their low toxicity and biocompatibility[23], [28]. Different plant extracts including the leaves, flowers, fruits, and roots have been used to synthesize AgNPs. For example, pomelo peel extract was used to synthesize AGNPs as reported by Nguyen [28]. In their work, they treated the peel extract with a citric acid solution for two hours at 85$^oC$, followed by mixing with silver nitrate solution ($AgNO_3$) and exposing the mixture to sunlight which induced the formation of 20-30nm AgNPs as characterized by transmission emission microscopy (TEM) [28]. Another study reports the use of the fruit extracts of the *Cleome Viscosa* plant to synthesize AgNPs which ranged between 20-50nm in size [29]. Polyphenols in rutin/curcumin have been reported in the synthesis of AgNPs via the reduction of silver ions by the carbonyl or phenolic hydroxyl groups of the rutin/curcumin [30]. Nanocellullose has also been used to synthesize AgNPs as reported by Yu and workers [31]. In their work, they reported the attraction of the positively charged silver ions by the negatively charged carboxyl and hydroxyl groups of the nano-cellulose leading to the reduction of the silver ions by the hydroxyl groups and themselves being oxidized to aldehyde groups [31]. Another research using corn starch to synthesize AgNPs has also been reported [32]. In this work, the authors revealed that the many hydroxyl and carboxyl groups on the linear chain of the starch were able to interact with the $Ag^+$ ions thus reducing them to stable AgNPs owed to the passivated surface of AgNPs and surrounded by negatively charged carboxyl groups [32]. More plant mediated synthesized AgNPs are summarized in table 2.

**Table 2: Plant-mediated synthesis of Silver Nanoparticles (AgNPs)**

| Plant | Plant part | Nanomaterials | Size (nm) | Shape | Reference |
|---|---|---|---|---|---|
| *peganum harmala* | Leaf extract | AgNPs | 20 | Spherical | [33] |
| Parsley (*Petroselinum crispum*) | Leaf extract | AgNPs | 30-32 | Spherical | [34] |
| *Ficus hispida* Linn. f. | Leaf extract | AgNPs | 20 | Spherical | [35] |
| *Phyllanthus pinnatus* | Stem extract | AgNPs | <100 | Cubicle | [36] |
| *Salvadora persica* | Root extract | AgNPs | 37.5 | Spherical and rod-like | [37] |
| *Alpinia nigra* | Fruit extract | AgNPs | 6 | Spherical | [38] |
| *Citrullus lanatus* | Fruit rind extract | AgNPs | 20 | Spherical | [39] |
| *Catharanthus roseus* | Bark extract | AgNPs | 1-26 | Spherical | [40] |
| *Cannabis sativa* | Stem fibre | AgNPs | 30 | Spherical | [41] |

### 2.3: Plant-mediated synthesis of Gold Nanoparticles (AuNPs)

Gold nanoparticles (AuNPs) have been reported to have good stability, size controllability, biocompatibility and, high adsorption capacity, and as such, have found wide applications in drug delivery, biosensors, gene delivery, water sanitation and, cancer therapy among others [23]. The green synthesis of the AuNPs has been done using the polyphenols, flavonoids, and amino acids from the extracts of leaves, flowers, roots, fruits, and seeds of various plants as reducing and stabilizing agents with tetrachloroauric acid (HAuCl4) as a precursor [23]. For example, Rodríguez-León and co-workers have reported the synthesis of 40nm AuNPs using the ethanol extract of *Mimosa tenuiflora* bark as the stabilizing and reducing agent of $HAuCl_4$ [42]. The AuNPs produced were found to be of good stability and less prone to agglomeration. Another study by Liu and co-workers reported the use of the polyphenols, flavonoids, and sugars of the leaf extract of *Cacumen platycladi* to synthesize AuNPs [43]. Likewise, Zangeneh and Zangeneh [44] have reported 15-45nm spherical AuNPs produced from the reduction of $HAuCl_4 \cdot 3H_2O$ using the water extracts of the *Hibiscus sabdariffa* flowers, that showed insignificant cytotoxicity to endothelial cells[44].

The polyphenols and anthocyanins in the fruits extracts of the Cornelian cherry (*Cornus mas*) plant have been reported to act as stabilizing agents in the synthesis AuNPs whose size ranged between 3-32nm with an average size of 15nm [45]. Also, the flavonoids in the plant extract of *Cannabis sativa* have been reported to reduce HAuCl4 to AuNPs with an average size of 25nm[46]. Similarly, the flavonoids rich *Alpinia nigra* leaves extract has been used to synthesize spherical AuNPs whose particles size was 21.52 nm and with photocatalytic activities [47]. Another study reports the synthesis of 33 nm spherical triangular and hexagonal AuNPs using the flavonoids in the leaf extract of the *Uncaria gambir Roxb*. as a bioreducing agent and triethanolamine as a capping agent [48]. Likewise, phytoconstituents isolated from the plant extract of *Trichosanthes kirilowii* have also been reported in the synthesis of 50 nm AuNPs which were found to inhibit the growth of colon cancer cells [49]. Many other plant-mediated syntheses of gold nanoparticles have been reported and some are summarized in table 3.

**Table 3: Plant-mediated synthesis of Gold Nanoparticles (AuNPs)**

| Plant | Plant part | Nanomaterials | Size (nm) | Shape | Reference |
|---|---|---|---|---|---|
| *Gnidia glauca* | Flower extract | AuNPs | 10 | Spherical | [50] |
| *Plumeria alba* | Flower extract | AuNPs | 15 - 30 | Spherical | [51] |
| *Moringa oleifera* | Petals extract | AuNPs | 3-5 | Spherical | [52] |
| *Garcinia mangostana* | Fruit peel extract | AuNPs | 32-38, 47 - 49 | Spherical | [53][54] |
| *Elettaria cardamomum* | Seed exctract | AuNPs | 15.2 | Spherical | [55] |
| *Lantana camara* | Berry exctract | AuNPs | 150–300 | Triangular | [56] |
| *Cannabis sativa* | Stem fibre | AuNPs | 20–40 | Spherical and triangular | [41] |
| *Capsicum annuum var. grossum* | Pulp extact | AuNPs | 6 - 37 | Trianglar, hexagonal, and quasi-spherical | [57] |
| *Artemisia vulgaris L.* | Leaf extract | AuNPs | 50-100 | Spherical, triangular and hexagonal | [58] |
| *Panax ginseng* | Leaf extract | AuNPs | 10 - 20 | Spehrical | [59] |
| *Trapa natans peel extract* | Peel extract | AuNPs | 26 - 90 | Spherical, triangular & hexagonal | [60] |

### 2.4: Plant-mediated synthesis of Copper Nanoparticles (CuNPs)

Copper nanoparticles (CuNPs) have exceptional properties for instance rigor, extraordinary yield strength, large surface area, elasticity, and ductility [61]. These properties make them have an extensive range of applications in catalysis, optics, antimicrobial activities as well as dye degradation [62], [63][64]. Normally, the CuNPs offer various advantages over traditional heterogenous catalysts hence increasing the environmental impact and economy [65]. Recently, they have gained a lot of reputation comparable with gold and silver on numerous grounds, since they are cheaper[61]. The synthesis of Cu-NPs using plant extracts has been commended as a valued substitute to both the physical and chemical methods since its ecofriendly, relatively economical, and involves an easy process [61]. Just like AgNPs and AuNPs, the phenolic compounds, flavonoids, and alkaloids found in plant extracts have been used as reducing and stabilizing agents in the synthesis of CuNPs. For example, Sajadi et al., [66] have reported the use of polyphenolic flavonolignans from *Silybum marianum* L. seed extracts as reducing agents to reduce copper ions in the synthesis of copper nanoparticles (CuNPS) transforming their structure from keto to enol [66]. Likewise, the leaf extracts of *Ginkgo biloba L.* have been used as reducing and stabilizing agents in the synthesis of CuNPS [65]. Another study reports the use of the biomolecules present in the leaves extracts of *Camelia sinensis* accountable for the reduction and stabilization of CuNPs [64]. The floral extracts of the *Quisqualis indica* have also been used as stabilizers in the synthesis of CuNPs [67]. Many other plants used in the synthesis of CuNPs include *P. granatum* seeds extract [68], *Allium noeanum* leaf extracts [69], *C. spinose* fruit extract [70], *Gymnema sylvestre* leaf extract [71], *Azadirachta indica* leaf extract[72], and seedless dates [73] among others.

### 2.5: Plant-mediated synthesis of Platinum Nanoparticles (PtNPs)

Platinum nanoparticles (PtNPs) have found wide applications in catalysis and biomedical fields due to their exceptional structural, optical and catalytic properties [74][75]. They have been reported to have antimicrobial, antioxidant, and anticancer properties[75]. Since the conventional chemical and physical synthetic methods for the PtNPs have been found to be harmful and costly, utilizing high energy, several researchers have utilized the readily accessible and abundant plants and plant extracts in the synthesis of PtNPs. The phytochemicals such as alkaloids, terpenoids, saponins, tannins, hydroxyls, and carbohydrate groups found in plants and plant extracts among others have been used as reducing and capping agents in the biosynthesis of PtNPs [74]. For example, the hydroxyl and carbohydrates groups found in the aqueous extract of the water hyacinth plant were used to efficiently reduce and stabilize PtNPs as reported by John Leo and coworkers [76]. Another study by Şahin and coworkers reports the biosynthesis of PtNPs from the plant extracts of *Punica granatum* crusts which were utilized as anti-tumor agents on the human breast cancer cells [77]. In yet another study, tea polyphenols have been reported as both reducing and surface modifying agents in the biosynthesis of PtNPs [78]. The protein biomolecules found in the Neem plant extract were used to reduce chloroplatinic ions into PtNps as reported by Thirumurugan and coworkers [79] in their research. Many other plant-mediated syntheses of PtNPs using plant extracts reported in the literature include black cumin seed (*Nigella sativa L.)* extract [80], Saudi's Dates extract [81], terrestrial weed *Antigonon leptopus* [82]*, Garcinia mangostana L.* fruit rind extract[82], *Coffea Arabica* seed extract [83], *Sechium edule* fruit extract [84], *Xanthium strumarium* leaf extract [85] among others.

**2.6: Plant-mediated synthesis of Carbon Nanotubes (CNTs)**

Carbon nanotubes (CNTs) are $sp^2$ hybridized nanocarbon materials composed of rolled-up graphene sheets [86] and are classified based on the number of walls as single-walled, double-walled, and multi-walled CNTs [87]. They are important nanomaterials because of their exceptional physical and chemical properties which include optical, electrical, and mechanical characteristics [87][88]. They have been utilized in sensors, solar fuels, supercapacitors membranes, and applied in biomedicine as biosensors [87]. Some of the methods that have been used in the synthesis of CNTs use metal catalysts which have been reported to cause severe threats to the living organisms as the catalysts linger in the CNTs and can be introduced in the human body in therapeutic applications [88]. Given the above-mentioned shortcoming of the conventional metal catalysts used, a green catalyst synthesized from the garden grass, rose, neem and wall-nut plant extracts in the synthesis of CNTs has been reported by Tripathi and coworkers [88]. Their results indicated that the catalysts prepared from the plant extracts led to the growth of multi-walled CNTs (MWCNTs) of ultra-long tubes of 8-15 nm in diameter.

## 3. Microbe Mediated Biosynthesis of Nanomaterials

The biological synthesis methods may take place intracellularly or extracellularly using biological agents such as bacteria, fungi, yeast, viruses, algae, or plants [89]. These methods present suitable alternatives to the physical and chemical methods of synthesis, owing to their ecofriendly procedures and the production of biocompatible non-toxic nanomaterials. Further, the metabolic profiles of the microbial makeup can be genetically engineered to allow for the fabrication of nanomaterials with specific biological, chemical, and physical properties [90]. These microbes can also adapt to diverse metal concentrations and possess the ability to reduce inorganic materials into nano particles via intracellular or extracellular pathways [91]. As a result, the biological synthesis of nanomaterials has attracted a lot of attention in the recent past and many authors have reported success in various syntheses schemes and bio agents [91], [92]. The most-reported microbial agents for nanomaterial synthesis are yeast, fungi, microalgae, bacteria, viruses, and actinobacteria [93]. Bacteria and microalgae give rise to well-defined nanomaterials such as nanowires, nano-cellulose, and exopolysaccharides [94], [95]. Different species of bacteria such as *Bacillus licheniformis, Pseudomonas deceptionensis, Pyrococcus furiosus, Pyrobaculum islandicum,* and *Pseudomonas aeruginosa*, among others, have been reported for the synthesis of various nanomaterials [96]. The section below discusses microbe-assisted mechanisms for nanomaterial synthesis.

### 3.1: Mechanisms for microbe assisted synthesis of nanomaterial

It has already been mentioned that microbe-assisted synthesis of nanomaterials may be intracellular or extracellular. Whether intracellular or extracellular, the nanomaterial formation involves absorption of ionic precursors by the microbes, followed by enzymatic reduction or oxidation into their respective nanomaterials [89]. This mechanism is direct for the synthesis of metal nanoparticles where the metallic ions are reduced into metallic nanoparticles.

In extracellular synthesis, the precursors are adsorbed outside the cell surface, where they undergo enzyme-catalyzed transformation into nanoparticles. The first step is usually the growth of the microbe in suitable media. This is followed by pre-concentration of the broth containing microbial cells via centrifuge [97]. A supernatant containing microbial enzymes is obtained for use in synthesis. For the synthesis of metal nanoparticles, the supernatant should contain reductase enzymes, which are allowed into contact with metal ions. The bioreduction takes place in a cell-free supernatant resulting in the formation of nanoparticles.

In intracellular synthesis, microbial cells absorb the precursors which then undergo enzyme-catalyzed transformation into nanoparticles within the cells. In the fabrication using the cellular mechanism of microbial cells, the first step involves the preparation of microbial cultures in suitable media. This is followed by washing the microbial biomass with sterile water, then centrifuging to obtain biomass pellet [98]. This microbial biomass is put in aqueous solutions of the desired metal ions and incubated. During incubation, the positively charged metal ions are trapped within the negatively charged cell walls of the microbial cells. While entrapped, the metal ions undergo enzyme-catalyzed bioreduction, resulting in nanoclusters, which diffuse into the solution [99]. The formation of the nanoparticles, in this case, is signified by the formation of specific colours that characterize particular nanoparticles. For instance, the synthesis of silver nanoparticles is signified by the appearance of pale yellow to brownish colour, while pale yellow to pinkish colour signifies the formation of gold nanoparticles [100]. The characterization of these nanoparticles usually requires techniques that can give the size, shape, crystallinity, and dispersity of the synthesized particles. Techniques such as transmission electron microscopy (TEM) and scanning electron microscopy (SEM), and X-ray diffraction spectroscopy (XRD) are used to estimate these characteristics. During synthesis, some proteins may be secreted and take part in capping the synthesized nanoparticles. Elucidation of the capping nature is usually examined using Fourier transform infrared spectroscopy (FTIR). The following section discusses various types of nanomaterials synthesized by microbial reactions.

### 3.2 Metal and metal oxide nanoparticles

Various metal nanoparticles have been synthesized via the reduction of metal ions to their respective metal nanoparticles. The most-reported microbes for the synthesis of metal nanoparticles include bacteria, actinobacteria, fungi, yeast, algae, and viruses. Metal nanoparticles synthesized by the use of microbes include AgNPs, AuNPs, CuNPs, SeNPs, and FeNPs, among others. For instance, silver nanoparticles are successfully produced via the reduction of silver ions to silver metal nanoparticles using different species of bacteria such as *Escherichia coli*, *Bacillus cereus*, *Acinetobacter* sp., *Pseudomonas* sp., and *Klebsiella pneumonia* among others. These different species of bacteria exhibit different rates of nanoparticle synthesis. Specifically, the species of *Escherichia coli, Klebsiella pneumoniae,* and *Enterobacter cloacae* have been shown to produce silver nanoparticles within five minutes of incubation [89], with *Enterobacter cloacae* showing the highest potency for speedy fabrication of the nanoparticles. On the other hand, the species *Bacillus licheniformis*, which contains nitrate reductase enzyme, has been shown to form silver nanoparticles within 24 hours of incubation. This underscores the importance of microbe choice in the synthesis process. Different sizes and shapes of silver nanoparticles are synthesized by changing the metal ions conditions during incubation such as culture medium, temperature, the concentration of the reactant, time of the reaction, and metallic salt among others [101]. This was demonstrated by Ramanathan et al., 2011 who used *Morganella psychrotolerans* species at different temperatures [102]. The authors observed the formation of hexagonal, triangular, and spherical nanoparticles at 25 °C. At 20 °C, spherical nanoparticles of 2-5 nm sizes were formed while a mixture of spherical and nanoplates was observed at temperatures between 15 and 20 °C. At low temperatures of 4 °C, spherical nanoparticles with 70-100 nm were formed. The effect of temperature, concentration, and pH on the biosynthesis of Ag nanoparticles was studied by Yumei et al, 2017 using Arthrobacter sp. B4 [103]. Using 1 mM concentration of $AgNO_3$, the authors obtained face-centered Ag nanoparticles whose size ranged between 9-72 nm at 70 °C and pH between 7-8. A further increase of temperature from 70 °C to 90 °C showed a significant decrease of synthesis time from 10 minutes to 2 minutes whereas increasing the concentration of $AgNO_3$ to 3 mM resulted in the formation of bunches of the Ag nanoparticles.

Gold nanoparticles are also another type of metal nanoparticle extensively synthesized using bacteria. Changes in the synthesis media affect an array of properties of these nanoparticles, just like it does for the Ag nanoparticles. For instance, Au nanoparticles with sizes of 10 - 20 nm were obtained using *Shewanella algae* at a pH of 7 but when the pH was adjusted to 1, various sizes of the synthesized nanoparticles ranged from 50 – 500 nm. In another report, spherical shaped Au nanoparticles with sizes of between 10 – 20 nm were obtained using

*Rhodopseudomonas capsulate* at a pH of 7 but the shapes became nanoplates when the pH was adjusted to 4 [104]. The authors further observed concentration-dependent shapes at pH 6, exemplified by the appearance of spherical shapes at low concentrations and nanowires at high concentrations. Studies have also established that the synthesis of Au nanoparticles using bacterial may involve enzymes or not. Non-enzymatic syntheses have been reported using dried cells of *Bacillus megaterium* [105], Corynebacterium sp. [106]. The enzymatic synthesis of nanoparticles is aided by cofactors such as NADH. This was used to obtain 20 – 25 nm-sized Au nanoparticles using Bacillus subtilis [107] and 5 – 50 nm sized particles using bacterium Geobacillus sp. strain ID17 [108]. NADH-dependent reductase enzymes from genetically engineered bacteria such as P. *Savastanoi*, R. *solanacearum*, P. *syringae*, and V. *fischeri* have also been used for the synthesis of Au nanoparticles [109]. Various sizes of nanoparticles between 20 nm and 40 nm have been achieved using these recombinant bacteria at neutral pH. Other types of metal nanoparticles recently synthesized using bacteria include 8 – 15 nm-sized Cu nanoparticles using *Pseudomonas stutzeri* and cubic Cu nanoparticles sized 50 nm – 150 nm using *P. stutzeri* [110]. Bacterial strains such as *Shewanella algae, Pantoea agglomerans, actobacillus acidophilus,* and *Azoarcus sp*. Have been shown to reduce selenite ions to selenium nanoparticles [98]

### 3.3: Actinobacteria mediated synthesis

Silver nanoparticles have also been synthesized using actinobacteria, which is a phylum under the bacterial domain and mainly consists of Gram-positive bacteria, commonly referred to as marine bacteria. Successful extracellular synthesis of Ag nanoparticles using these types of bacteria has been demonstrated by reduction of $Ag^+$ to Ag nanoparticles via NADH-depended nitrate reductase from the marine bacterium Streptomyces sp. LK-3 [111]. Other marine bacteria used in the synthesis of Ag nanoparticles include *Streptomyces rochei* MHM13 [112], *Streptacidiphilus durhamensis* [93], *Streptomyces xinghaiensis OF1* [113], *Streptomyces sp*. OSIP1, and *Streptomyces sp.* OSNP14 [114]. Other actinobacteria synthesized metal nanoparticles are those of gold and copper. Ranjitha and Rai (2017) were the first to report biosynthesis of Au nanoparticles using soil-isolated *Streptomyces griseoruber* actinobacteria and obtained 5 – 50 nm-sized Au nanoparticles [115]. Copper nanoparticles were synthesized using *Streptomyces capillispiralis* Ca-1 [116]. The authors obtained particle sizes of 3.6 – 59 nm using TEM analysis. Further, the authors established through FTIR that the actinobacteria contained bioactive functional groups that stabilize the Cu nanoparticles.

### 3.4 Fungi mediated synthesis

The synthesis of Ag nanoparticles has also been successfully achieved using fungi. This synthesis is reported to be more efficient and cheap, owing to fungi's high accumulation potential to metals. Moreover, the extracellular synthesis of nanoparticles using fungi is easily achieved since it does not involve doping as is required in intracellular synthesis. This method also does not need extra reagents such as detergents of modulation of physical factors such as ultrasound. Metal ions are known to highly bind to the cell walls of fungi and can accommodate high concentrations of metals. This allows higher yields of nanoparticles compared to those synthesized using bacteria. Successful extracellular synthesis of Ag nanoparticles has thus been demonstrated by Metuku et al., 2014 who obtained 10 – 40 nm-sized nanoparticles using *Schizophyllum radiatum* fungi [117]. Studies of the effect of pH on the fungi-mediated synthesis of Ag nanoparticles have shown that the shape of the nanoparticles can be modulated using pH changed. This is because pH changes determine the basicity or acidity of amino acids, which are involved in nanoparticle synthesis. This was demonstrated by Rajput et al, 2016 who obtained spherical, triangular, rod-shaped, and irregular shapes of Ag nanoparticles at a pH of 3 using *Fusarium oxysporum* 405 fungi [118]. The authors further observed the formation of monodispersed spherical nanoparticles at pH of 5 and 7 while pH of 9 managed the formation of long and spherical nanoparticles. On temperature studies, it has been established that *Fusarium oxysporum* mediated synthesis of Ag nanoparticles is optimum at 40 °C - 60 °C temperature range [119]. For *Fusarium oxysporum* 405 mediated synthesis of Ag nanoparticles, a temperature range of 50 °C – 70 °C has been found to produce maximum but smaller sized (10 nm) nanoparticles while 25 °C produces minimum but large-sized (50 nm) nanoparticles [118].

Gold nanoparticles are other materials synthesized using fungi and significant successes have been achieved. Kitching et al., 2016 used *Rhizopus oryzae* to synthesize 16 – 19 nm-sized Au nanoparticles [120]. The effect of temperature, incubation time, salt concentration, pH on the size and shape of the synthesized Au nanoparticles has also been studied. The studies have established that the rate of synthesis of Au nanoparticles increases with an increase in the temperature, incubation time, and salt concentration using *Pleurotus ostreatus*. Highly stable, 40 – 45 nm-sized Au nanoparticles were synthesized using *Fusarium solani* ATLOY-8 endophytic fungi [121], and 5 -10 nm-sized Au nanoparticles using *Cladosporium* sp. [122].

Some metal oxides have been synthesized using fungi. These include ZnO, CuO, CoO, $TiO_2$, as well as ZnO/CuO nanocomposite. ZnO nanoparticles of a size range between 10 45 nm were synthesized extracellularly using *Aspergillus terreus* AF1. The fungi were found to reduce bulk ZnO to ZnO nanoparticles, and the nanoparticles were capped by proteins secreted by the same fungi, as observed using FTIR [123]. The morphology of ZnO nanoparticles can be varied by changing the fungal strains used in synthesis. This was demonstrated by the synthesis of ZnO using soil-isolated *Fusarium keratoplasticum* A1-3 and *Aspergillus niger G3-1 strains* [124]. *The Fusarium keratoplasticum* A1-3 mediated ZnO nanoparticles were hexagonal and had a size range between 10 -42 nm while those mediated by *Aspergillus niger G3-1were nanorods sized 8 – 32 nm. In a different report, Xylaria acuta* mediated ZnO nanoparticles were also hexagonal and had a size range between 34 – 55 nm [125]. CuO nanoparticles were synthesized using *Penicillium chrysogenum* and gave nanoparticles of size 10.5 – 59.7 nm [126]. CuO was also synthesized using *Aspergillus niger G3-1 and characterized with SEM, TEM, XRD, and FTIR* [127]. *The authors obtained nanoparticles sized between 14 – 47.4 nm. A nanocomposite of CuO and ZnO was synthesized using Penicillium corylophilum* As-1, giving rise to nanoparticles of size 10 – 55 nm. $TiO_2$ was synthesized using *Trichoderma viride, where spherical nanoparticles with sizes between 60 – 86.67 nm were obtained* [128]*, and Pleurotus djamor* [129]*, giving 31 nm sized particles. Successful biosynthesis of iron oxide nanoparticles using fungi such as Fusarium incarnatum*, *Trichoderma asperellum*, and *Phialemoniopsis ocularis* is also reported [130], while CoO nanoparticles can be synthesized using *Aspergillus nidulans* [131].

### 3.5 Yeast mediated synthesis

Yeats, like fungi, can withstand high metal ion concentrations, thereby allowing high deposits of metal nanoparticles [10]. Biosynthesis of Ag nanoparticles using yeast has been demonstrated by Apte et al., 2013, who used a brown pigment from yeast, known as melanin from Yarrowia lipolytica [132]. In another report, successful extracellular biosynthesis of 20 -80 nm-sized Ag nanoparticles was achieved using *Candida utilis* NCIM 3469 yeast [100]. Smaller sizes of Ag nanoparticles (2 – 10 nm) were obtained in another report using a termite gut isolate *Candida lusitaniae* [133]. Other yeast-mediated syntheses of nanoparticles include hexagon, spheres, and triangles shaped Au nanoparticles using *Magnusiomyces ingens* LH-F1 yeast [134] and Pd nanoparticles using *Saccharomyces cerevisiae* extract [135]. Genetically engineered yeasts have been used to fabricate uniform sizes of nanoparticles. For instance, the fusion of *Pichia pastoris* yeast strain with a metal resistant gene from *Mucor racemosus* has been found to produce cytochrome b5 reductase enzyme [136]. This enzyme reduces metal ions and produces stable and uniform sized Ag nanoparticles. $TiO_2$ nanoparticles were synthesized using baker's yeast, and gave particle size of 6.7 nm.

### 3.6 Algae mediated synthesis

Algae is another unique bioagent used for the synthesis of nanoparticles. Its uniqueness arises from the fact that its cells have secondary metabolites and other bioactive compounds which cap the synthesized nanoparticles and stabilize them [137]. Highly stable -25 nm-sized Ag nanoparticles have thus been obtained using marine alga extracted *Caulerpa racemose* [138]. In another report, *Laminaria japonica* was used for biosynthesis of Ag nanoparticles and showed a significant rate of synthesis with incubation of $AgNO_3$ at 120 °C [139]. Spherical Ag nanoparticles were obtained using *Portieria hornemannii* red alga [140], while 25 – 60 nm-sized Ag nanoparticles were obtained using marine macroalgae *Padina* sp. [141]. Au nanoparticles have also been synthesized using algae. These include 5 – 35 nm-sized Au nanoparticles synthesized using *Tetraselmisko chinensis*, 53 – 67 nm-sized Au nanoparticles using *Padina gymnospora* [142], as well as round, hexagons, pentagons, triangles, and truncated triangle-shaped nanoparticles using *Euglena gracilis* [143]. Celluar extracts of *Turbinaria conoides* and *Sargassum tenerrimum* brown alga have also been used to synthesize 27 – 35 sized Au nanoparticles [144] while stable and spherical 8.4 nm sizes were obtained using *Cystoseira baccata* brown microalga [145]. Of these algae, the *Euglena gracilis* has been found to exhibit improved growth kinetics, high colloidal stability, high yields, and high growth rate [143]. Palladium nanoparticles using have also been synthesized using algae species such as *Sargassum Bovinum* [146]*, Chlorella vulgaris* [147], and *Spirulina platensis* [148]. Silver chloride nanoparticles present unique metal salt nanoparticles that have been synthesized using algae. Examples are 21 – 48 nm AgCl nanoparticles synthesized using *Sargassum plagiophyllum* [149], 25 nm-sized AgCl nanoparticles using *Caulerpa racemose* [138], and 9.8 nm-sized AgCl nanoparticles using *Chlorella vulgaris* [150].

Metal oxide nanoparticles have also been synthesized using algae. For instance, zinc oxide nanoparticles have been synthesized extracellularly using *Turbinaria conoides* and *Padina tetrastromatica*, both marine algae; *Sargassum muticum* [92], extract from green microalgae *Chlorella* [151], blue-green algae *Arthrospira platensis* [152]*,* and an extract from green algae *Chlorella vulgaris* [153]. These nanoparticles were characterized using SEM and TEM, FTIR, and XRD. The green microalgae Chlorella synthesized nanoparticles were 20 nm in size, monodispersed, and hexagonal, the *Arthrospira platensis* synthesized nanoparticles were spherical with sizes of 30 – 55 nm, while those synthesized using *Chlorella vulgaris* were nanorods in shape with lengths of 150 nm and width of 21 nm. In another study, $TiO_2$ nanoparticles and $TiO_2$-graphene oxide nanocomposites were synthesized using green algae *Chlorella pyrenoidosa.* When observed through XRD and TEM, the $TiO_2$ nanoparticles showed spherical shapes sized 50 nm while the $TiO_2$-graphene oxide nanocomposites formed sheet-like structures [154]. Iron oxide nanoparticles *Sargassum acinarium* and *Padina pavonica* brown algae seaweeds [155]. TEM analyses of these nanoparticles showed that particles sizes of between 21.6 – 27.4 nm were obtained using *Sargassum acinarium* while 10 – 19.5 nm were obtained using *Padina pavonica*.

### 3.7 Virus mediated synthesis

The virus-mediated biosynthesis of nanoparticles benefits from nanoscale capsid proteins that cover the viruses, and which provide effective interactions with metallic ions [89]. The capsid proteins also allow for modification through genetic engineering, which facilitates the synthesis of nanocomposite and nanoconjugate metallic particles [156]. The most preferred viruses for nanoparticle synthesis are those extracted from plants. This is because these viruses are nonpathogenic to humans and animals, are stable, and are easy to cultivate. They are therefore suitable for the synthesis of nanoparticles whose applications are in biomedical studies. Viruses have therefore been used for the synthesis of both mono-metallic and bimetallic nanoparticles. Mono-metallic Au nanoparticles have been synthesized using cowpea chlorotic mottle viruses [157] while uniform 5 nm-sized Au nanoparticles were synthesized using tobacco mosaic viruses [158]. Bi-metallic nanoparticles of Au-Ag were synthesized using pathogenic squash leaf curl China virus as a bio template [159]. The obtained nanoparticles were electroactive and biocompatible. A few metal salt nanoparticles have been synthesized using viruses. These include the fabrication of CdS and ZnS using M13 bacteriophage [160]. The above few reports suggest that the use of viruses for the synthesis of nanoparticles has not been extensively explored. This is attributed to encountered challenges such as lack of large-scale applicability, limited biocompatibility, and the special requirements for virus-mediated synthesize of the nanoparticles.

## 4. Biomedical Application of green synthesized nanomaterials in Biosensors

Biosensors can be defined as self-sufficient analytical devices capable of transforming a biological response into a measurable and processable analytical signal [161]. The biosensor consists of a biological recognition element such as nucleic acid, enzymes, cells, or tissues to especially react with the analyte, a physical transducer such as electrochemical, optical, thermal, or piezoelectric, which transform the analyte-based receptors into the output signals, and a signal output display [162]. They are therefore classified based on the transducer or the biorecognition element [163]. They have found a wide range of applications in the biomedical field for medical diagnostics because they are easy, fast, inexpensive, very sensitive, and highly selective [164]. Nanomaterials, due to their exceptional properties, have been used as transducer materials in biosensor development which has led to enhanced analytical signals and improved biosensors performance including enhanced sensitivities and selectivities [165]. Green synthesized nanomaterials play a vital role in the development of environmentally friendly biosensors. Various biomolecules including enzymes, amino acids, proteins, lipids, and DNA, and ions which are found in physiological fluids such as blood, saliva, and urine have been reported to provide a hostile environment for nanomaterials such as AuNPs and AgNPs [162]. This can however be overcome by the choice of the capping agents and the thickness of the layer as well as the surface chemistry as important factors which must be taken care of when fabricating the nanobiosensors [162]. The biosensing mechanism in nanobiosensors is based on the modulation of the nanomaterials signals in response to the analyte interactions with the capping agents and surface molecules that govern the distance-dependent interactions across the capping layer. This in turn induces resonance energy transfer interactions generating DNA-DNA hybridization, varying the effective charge transfer, and stimulating the fluorescence as illustrated in figure 4 using Au as a representative.

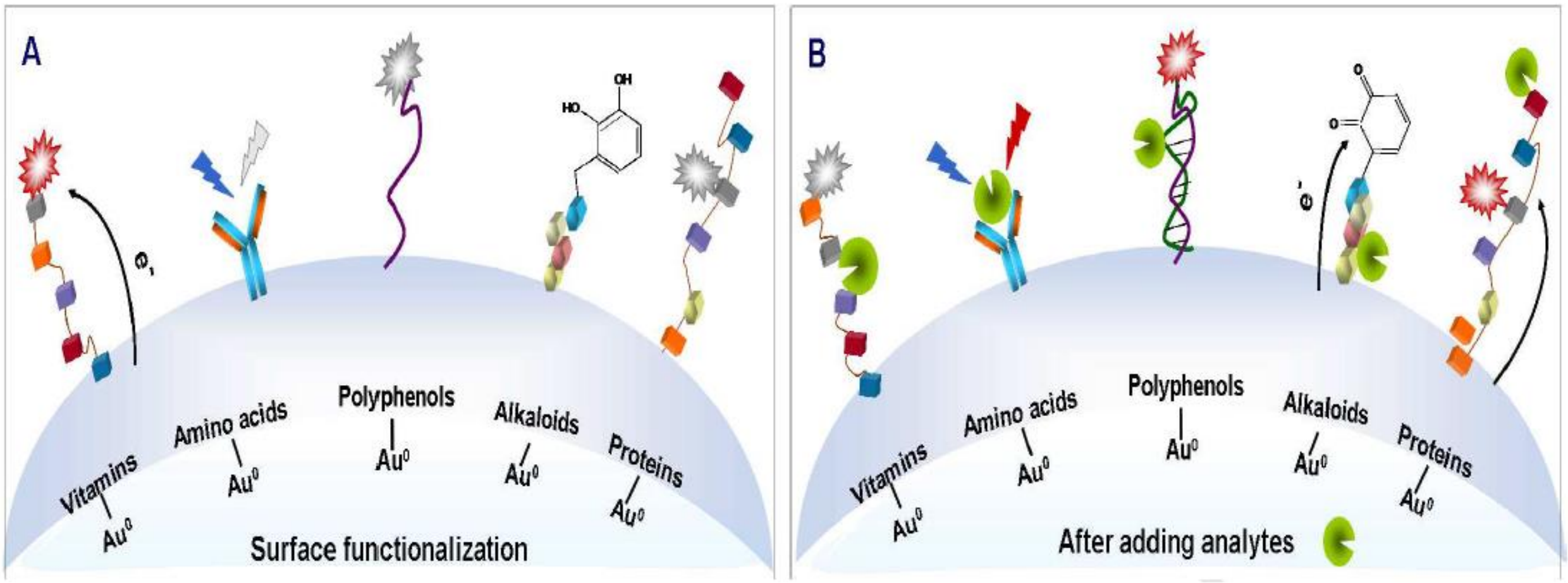

Figure 4: A schematic illustration of the biosensing mechanism in nanobiosensors. (A) The optical surface-bound biomolecular sensing structures of the AuNPs and (B) their biosensing mechanisms are based on charge transfer, fluorescence quenching, resonance energy transfer, DNA-DNA hybridization, and analyte-biomolecule interactions. Reproduced with permission from [162]. © 2020 Elsevier B.V. All rights reserved.

As described in section 2, various nanomaterials have been used in biosensing due to their low toxicity and biocompatibility and their exceptional properties. For example, as reported in the literature by Bollella *et al.,* [166] AgNPs, synthesized using quercetin as the reducing agent was used to modify graphite electrodes which were later used to fabricate a third-generation lactose biosensor characterized by a good linear range, enhanced sensitivity due to low detection limits, good stability and rapid response time [166]. Likewise, AgNps synthesized from the onion peel have also been used in the development of a biosensor for the detection of toxic mercury in the liquid phase [167]. Green synthesized AgNPs have also been used in the development of biosensors for the detection of drugs as reported by Zamarchi and Vieira [168]. In their work, they used AgNPs synthesized using pine nut extract (*Araucaria angustifolia*) as a reducing and stabilizing agent to develop an electrochemical sensor for the determination of paracetamol whose performance was symbolized by a good linear range with a detection limit of 8.50 x $10^{-8}$M and good stability. This indicated that the sensor could be used to quantify paracetamol in pharmaceutical products [168].

Green synthesized AuNPs have also been used in biosensors. For example, chitosan synthesized AuNPs were conjugated to an antibody and used to develop an optical biosensor for a target antigen [169]. Also, fluorescent probes from AuNPs synthesized using papaya juice as the capping and reducing agents were used to develop a highly selective, biocompatible, and stable biosensor for the detection of L-lysine [170]. The authors of this work reported a fluorescence intensity that was linearly dependent on the concentration of L-lysine in the range of 10.0 µmol/L to 1000.0 µmol/L and a detection limit of 6.0 µmol/L. They further used these biosensors to monitor L-lysine in urine with their results demonstrating great potential in use in real samples [170]. Another study reports the use of gold nanoclusters synthesized with the protein from soybeans as capping agents to develop a probe for the detection of bismerthiazol with a response range of 5 – 1000 µg/mL, which was attributed to the ability of the bismerthiazol to interact with the gold nanoclusters [171]. In yet another study, gold nanoclusters synthesized with the soybean extract as the capping agent were used to develop a sensor for copper ions with a god sensitivity and selectivity and showed a great application potential in food analysis and the biomedical field [172]. Papain-stabilized gold nanoclusters with a red fluorescence emission at 639 nm were used in the development of a biosensor for sensitive and selective detection of D-penicillamine as reported by Chen and coworkers [173]. Their sensor was linear in the range of 30.0 µM to 2.0 mM and a limit of detection of 5.0 µM and was further utilized in exploring the metabolic process of the D-penicillamine in rats indicating a potential application in clinical studies [173]. Another study also reports the use of gold nanoclusters synthesized from onion membranes in the fabrication of sucrose sensor [174].

Biosynthesized copper nanoparticles (CuNPs) have also been utilized in the development of biosensors. As reported by Dayakr and coworkers, CuNPs synthesized using *ocimum tenuiflorum* leaf extract were utilized to develop a nonenzymatic electrochemical-based glucose biosensor [175] which was prepared by coating the CuNPs on a glass carbon electrode and studying the electrochemical response using cyclic voltammetry and amperometry. The biosensor demonstrated a remarkable selectivity, sensitivity, and a prompt response time of less than 3s and a wide linear range of 1 – 7mM and a detection range of 0.038 µM [175]. Copper oxide nanoparticles (CuONPs) synthesized using *Caesalpinia bonducella* seed extract as reducing, stabilizing, and capping agents have also been utilized in the construction of an electrochemical biosensor for the detection of riboflavin [vitamin $B_2$ (VB2)] [176]. The electrochemical sensor made from a modified electrode demonstrated good stability, reproducibility and could be used for nanomolar detection of VB2 at a linear range of 3.13–56.3 nM and a detection limit of 1.04 nM [176]. Likewise, CuONPs synthesized from the stem latex of peepal (*Ficus Religiosa*) have been utilized in the fabrication of an electrochemical biosensor for the detection of pesticides [177].

## 5. Biomedical Application of green synthesized nanomaterials in Drug Delivery

Nanoparticles have a characteristic large surface area to volume ratio, resulting from their small sizes. This property makes them access and penetrates cell walls, thereby allowing the possibility of delivering drugs to targeted sites of the body. For this reason, various types of nanoparticles have been synthesized, modified, and applied as drug carriers across the body.

This has made it possible to detect various diseases at onset and initiate early treatment. Nanoparticles have found applications in medicine as either agents against disease-causing pathogens or nanocarriers of drugs to targeted sites of infection. This section discusses the applications of biosynthesized nanoparticles in drug delivery.

Nanoparticles that have exhibited actions against diseases include platinum, selenium, silver, and gold. Platinum nanoparticles synthesized using *Saccharomyces boulardii* were found to exhibit anticancer properties against MCF-7 and A431 cell lines [178]. Silver nanoparticles synthesized using *Cryptococcus laurentii* were reported to have anticancer properties against breast cancer cell lines [179]. Selenium nanoparticles obtained from extracellular synthesis using *Streptomyces bikiniensis* were found to initiate mortality of Hep-G2 and MCF-7 cancer cells [180]. In a different report, selenium nanoparticles synthesized using *Streptomyces minutiscleroticus* M10A62 inhibited the growth of HepG2 and HeLa cell lines [181]. The use of nanoparticles for the treatment of diseases is usually met with challenges such as toxicity to non-targeted areas, dosage, and host immune responses. These challenges have been significantly addressed by encapsulating drugs in nanocarriers and for delivery to specific sites of infection. To demonstrate success in this, zinc oxide nanoparticles were synthesized using *Rhodococcus pyridinivorans* and used as nanocarriers for anthraquinone. [182]. The anthraquinone-loaded nanoparticles were studied for the treatment of HT-29 colon carcinoma cells. The nano-delivered drug was found to exhibit cytotoxicity against the colon carcinoma cells. Biosynthesized gold nanoparticles have also been used to deliver doxorubicin drugs and showed a high diffusion rate into KEK293 cancer lines [183], and hepatic cancer cells[89]. In another report, gadolinium oxide nanoparticles have shown effective delivery of taxol drugs for the treatment of cancers [184].

## 6. Challenges of green synthesis of nanomaterials

The research on the biosynthesis of nanoparticles and their applications has witnessed tremendous successes but has also been met with several challenges. The nanoparticles discussed in this work are majorly metallic such as silver, palladium, and gold, as well as metallic oxides of copper, zinc, cobalt, titanium, and iron. The growing research in these materials and their applications has been accompanied by the release of these nanoparticle residues to the environment. This presents one of the major pathways through which these materials enter into the environment, the food chain, and consequently alter the ecological balance. Toxicology studies of the nanoparticles have shown that they cause DNA damage, alter cell membranes and generate reactive oxygen species [185]. The metal nanoparticles highly adsorb on cell walls, depolarizing them and permeating into the membranes where they cause disintegration. The reactive oxygen species can target and bind to multiple sites of the cell at the same time. This causes conformational changes in the proteins, lipids get peroxidated and the DNA is damaged. The consequence of this is that the membrane is disintegrated and the cell dies [186]. Silver nanoparticles synthesized from *Sargassum siliquosum* have been shown to increase the levels of serum creatinine and blood urea nitrogen [187] while those synthesized using *Ficus religiosa* leaf extract accumulate in the brain, liver, and lungs. Similarly, Au nanoparticles synthesized using *Helianthus tuberosus* accumulate in the liver, lungs, kidney, and spleen. It is therefore a potential cause of organ damage, especially the lungs, which are the main target [188].

Another challenge is the biocompatibility of the metal and metal oxide nanoparticles. This calls for further modification of these particles so that biocompatible groups are introduced. The nanoparticles are also unstable in a wide variety of matrix mixtures, especially in physiological environments, which are desired for biomedical applications. Surface capping is, therefore, necessary to improve the stability of the nanoparticles. These extra synthesis procedures are associated with increased time and cost of synthesis. Furthermore, the increased layers during functionalization or capping may increase the size of the particles and often interfere with the desired properties of the synthesized nanoparticles.

## 7. Conclusions and Future perspectives

This review has given comprehensive recent trends in the green synthesis of nanoparticles and their biomedical applications. The focus was given on plant-mediated as well as the microbe-mediated synthesis of nanoparticles. Significant advances have so far been made in the green synthesis of nanoparticles and their applications in drug delivery. It is noted from the review that plant-mediated synthesis is more embraced compared to microbe-mediated synthesis. Furthermore, virus-mediated synthesis has attracted very few publications, an indication that this method has not been widely explored. Future focus on virus-mediated synthesis should be on large-scale applicability, increased biocompatibility, and the innovations of alternative approaches to the special requirements for virus-mediated synthesize of the nanoparticles

It is widely acknowledged by all authors that green synthesis is a suitable alternative to other methods of synthesis as it reduces the use of harsh chemicals and releases hazardous wastes to the environment. It is no doubt that the ecofriendly green synthesis should extend not only to nanoparticles but also to all other syntheses in the future. However, for nanoparticles particularly, their widely practiced synthesis and applications have led to the release of nano residues to the environment, at a time when the toxicology studies of these particles and disposal are not fully understood. There are reports of the accumulation of nanoparticles on vital body organs, usually causing cell death and organ damage. Future research needs to focus on effective means through which these nanoparticle residues are eliminated from the system without compromising the time required for delivery of adequate dosage treatment. Single-step synthetic routes have been explored whereby nanoparticles are synthesized, stabilized, and functionalized in readiness for bioconjugation. This presents a significant achievement in the fabrication of nanoparticle bioconjugates. It is however desirable to ensure that the capping and functional groups added do not alter the properties of the nanoparticles.

## 8. Acknowledgments

The authors would like to thank the School of Pharmacy and Health Sciences - United States International University-Africa and the Technical University of Kenya for providing a conducive environment and facilities from which this work was written.